\documentclass[runningheads]{cl2emult}

\usepackage{makeidx}  % allows index generation
\usepackage{graphicx} % standard LaTeX graphics tool
\usepackage{subeqnar} % subnumbers individual equations
\usepackage{multicol} % used for the two-column index
\usepackage{cropmark} % cropmarks for pages without
\usepackage{eso}      % placeholder for figures
\makeindex            % used for the subject index
\def\ls{{_<\atop^{\sim}}}
\def\gs{{_>\atop^{\sim}}}
\def\cgs{ ${\rm erg~cm}^{-2}~{\rm s}^{-1}$ }

\begin{document}
\title*{High Energy Large Area Surveys:\protect\newline from BeppoSAX to
Chandra and XMM}
\toctitle{High Energy Large Area Surveys:\protect\newline from BeppoSAX to
Chandra and XMM}
% allows explicit linebreak for the table of content
%
%
\titlerunning{HELLAS surveys}
% allows abbreviation of title, if the full title is too long
% to fit in the running head
%
\author{Fabrizio Fiore\inst{1}
\and Andrea Comastri\inst{2}
\and Fabio La Franca\inst{3}
\and Cristian Vignali\inst{4}
\and Giorgio Matt\inst{3}
\and G. Cesare Perola\inst{3}
\and the HELLAS collaboration}
\authorrunning{Fabrizio Fiore et al.}
% if there are more than two authors,
% please abbreviate author list for running head
%
%
\institute{Osservatorio Astronomico di Roma, via Frascati 33, 
Monteporzio, I00040, Italy
\and Osservatorio Astronomico di Bologna, via Ranzani 1, I40127
Bologna, Italy
\and Dipartimento di Fisica, Universit\`a Roma Tre, 
Via della Vasca Navale 84, I--00146 Roma, Italy
\and Dipartimento di Astronomia, Universit\`a di Bologna, via Ranzani 1, 
I40127 Bologna, Italy
}

\maketitle              % typesets the title of the contribution

\begin{abstract}

Optical identification of hard X-ray selected BeppoSAX and Chandra
sources indicates that a large fraction of the sources are
``intermediate'' AGN, i.e. type 1.8-1.9 AGN, broad-line quasars and
even X-ray loud but optically dull (apparently normal) galaxies, all
obscured in X-rays by columns of the order of $10^{22-23.5}$
cm$^{-2}$.  Because of this obscuring matter, these sources 
are more difficult to detect or select at other wavelengths,
implying that a fraction of the accretion power in the Universe may 
have been missed so far.

\end{abstract}

\section{Introduction}
The characteristic shape of the X-ray Cosmic background (XRB) strongly
suggests that most (80-90 \%) of the accretion luminosity in the
Universe is obscured (e.g. Hasinger et al. 1999, Fabian \& Iwasawa
1999, Hasinger et al. 2000).  Hard X-ray surveys are therefore the
most efficient way to trace accretion, since obscured, accreting
sources are more difficult to detect or select at other
wavelengths. Furthermore, the source classification based on optical-UV
lines is based on ``secondary'' properties (e.g. emission lines from
ionized plasma) and so may be inaccurate and/or incomplete (see for
example the cases of NGC6240, Vignati et al. 1999 and NGC4945,
Guainazzi et al. 2000).  Conversely, hard X-ray selection and
classification are based on ``primary" properties: the emission from a
region close (10-100 gravitational radii) to the central supermassive
black hole.  Hard X-ray surveys can then be used to record the
``accretion history'' of the Universe, i.e. the history of the light
ultimately produced by gravity.  This can be compared with the history
of the formation of structures in the Universe, from proto-galaxies,
to galaxies, to groups and clusters, which again is driven by gravity,
and to the history of the star-formation, i.e.  the history of the
light produced by nuclear reactions in stars.  These comparisons can
give us a clue on the correlations between the formation and light-up
of supermassive black holes in galactic nuclei and galaxy formation
and evolution. For example, Fabian (1999) has proposed a scenario in
which powerful radiation driven winds emerge from newly born quasars,
possibly affecting the star-formation processes in the nearby gas
clouds and the mixing of the ISM in the host galaxy.  In this scenario
the powerful high redshift, newly born quasars are expected to be
highly obscured (actually the majority of them would be
Compton-thick), and therefore nearly invisible at optical-UV and soft
X-ray energies.  Alternatively, the bulk of the hard XRB may be due to
a large population of relatively low luminosity (Seyfert-like) AGN at
moderate redshift (z$<1$). If this is the case a large fraction of the
galaxies at these redshifts should host an active nucleus.  Hard X-ray
surveys can help in disentangling between these competing
scenarios. We present here a work-in-progress on our High Energy Large
Area Survey which makes use at the moment of BeppoSAX and Chandra
data. XMM data will be included as soon as the first fields will
become public.  Our approach is complementary to deep pencil beam
surveys (e.g. Giacconi et al. 2000, Hasinger et al. 2000) in that we
cover different portions of the redshift--luminosity plane.  Our
purpose is to study cosmic source populations at fluxes where a large
fraction of the hard XRB is resolved ($\approx50\%$), but where a) the
X-ray flux is high enough to provide at least rough X-ray spectral
information; b) the area covered is as large as possible, to be able
to find sizeable samples of ``rare'' objects; and c) the magnitude of
the optical counterparts is bright enough to allow relatively high
quality optical spectroscopy, useful to investigate the physics of the
sources.

\section{X-ray data and optical identifications}

The BeppoSAX HELLAS survey covers $\sim55$ deg$^2$ of the sky with
$\delta<+79$, $20<\alpha<5$ and $6.5<\alpha<17$ at 5-10 keV fluxes in
the range $5\times10^{-14}-10^{-12}$ \cgs (Fiore et al. 2000a). The
Chandra survey covers about half deg$^2$ at a 2-10 keV flux limit of
$10^{-14}$ \cgs (Fiore et al. 2000b, Cappi et al. 2000).  About half
of the BeppoSAX (62) and Chandra (17) sources have been optically
identified. Because of the quite large MECS error box ($\sim1'$
radius, Fiore et al. 2000a), we limit the optical identification
process of BeppoSAX sources to objects with surface density $\ls40$
deg$^{-2}$, to keep the number of spurious identifications in the
whole sample smaller than a few percent. This translate in the
following limits on the magnitude of the possible optical
counterparts: R$<20.5$ for broad-line AGN; R$<19$ for narrow-line AGN;
R$<17.5$ for emission line galaxies (LINERs and starburst
galaxies). According to these limits about one third of the BeppoSAX
error boxes studied in detail remain ``unidentified''.  Results on the
optical identification of BeppoSAX and Chandra hard X-ray sources have
been presented by Fiore et. al. (1999), Fiore et al. (2000b), Cappi et
al.  (2000), La Franca et al. (2001), in preparation. Here we limit the
discussion to the following three main topics.

\subsection{The fraction of obscured to unobscured AGN}

To study the spectral variety of the HELLAS sources we have calculated
for each source the softness ratio (S-H)/(S+H) (S=1.3-4.5 keV,
H=4.5-10 keV count rates for the BeppoSAX sources and S=0.5-2 keV,
H=2-10 keV fluxes for the Chandra sources).  (S-H)/(S+H) is plotted as
a function of the redshift in figures 1a (BeppoSAX) and 1b
(Chandra). While broad line AGN are identified up to z=2.76, all
narrow line AGN in figure 1a have z$<$0.4. This is due very likely to
the conservative threshold adopted for the optical magnitude of narrow
emission line AGN and galaxies (which do not show a bright optical
nucleus, because of the strong extiction).  Because of its relatively
large error-boxes BeppoSAX cannot be used to unambiguously identify
high redshift, narrow line AGN and galaxies.  The dotted lines in
figures 1a,b represent the expectation of unabsorbed power laws with
$\alpha_E=0.4$ and 0.8.  The dashed lines represent the expectations
of a power law model absorbed by columns of increasing densities
(in the source frame). Many BeppoSAX and Chandra HELLAS sources have
(S-H)/(S+H) inconsistent with that expected from unabsorbed power law
models with $\alpha\sim0.8$. Absorbing columns of the order of
$10^{22.5-23.5}$ cm$^{-2}$ are most likely implied. The fraction of
highly obscured sources ($N_H\gs10^{23}$) in figure 1a at z$<0.3$
(where our survey should be representative of the actual source
population) is $\sim0.42$. This is consistent with the expectations of
AGN synthesis models (Comastri et al. 2000 in preparation). In this
redshift range all highly obscured objects are narrow line AGN and
galaxies, while all broad line AGN have $N_H<10^{23}$ cm$^{-2}$,
consistent with popular AGN unification schemes.
 
\begin{figure}
\centering
\hbox{
\includegraphics[width=.52\textwidth]{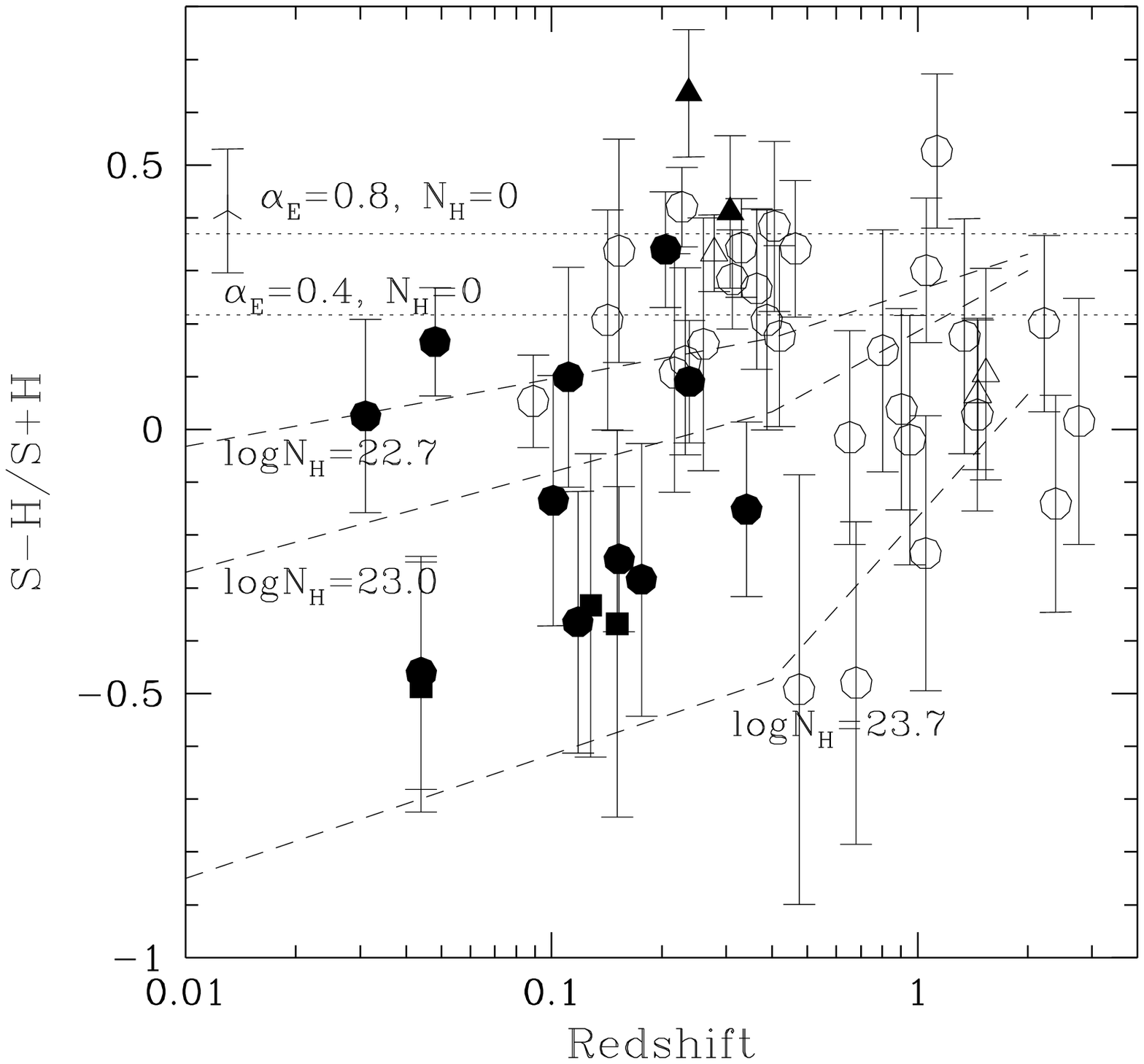}
\includegraphics[width=.52\textwidth]{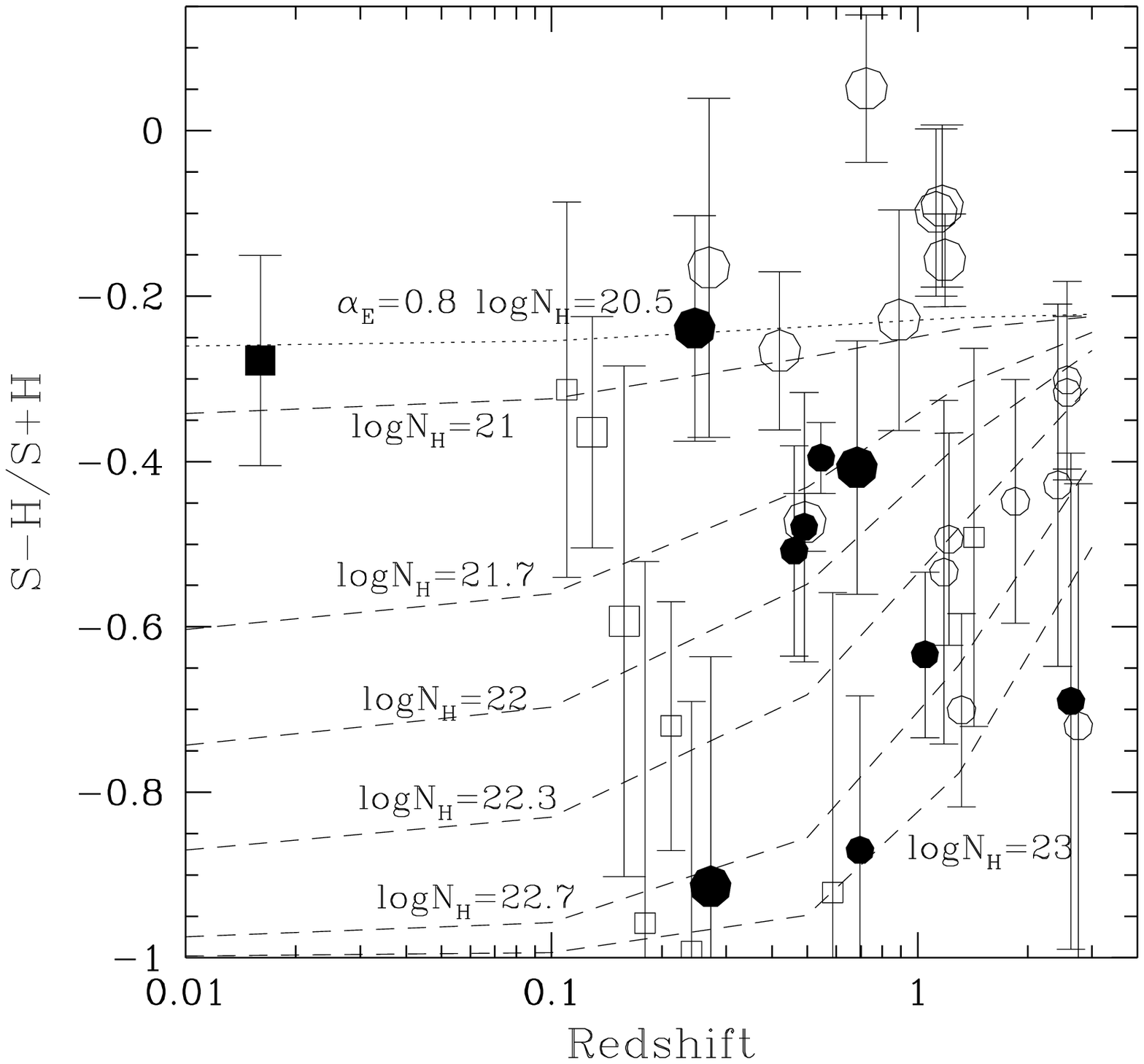}
}
\caption[]{a) BeppoSAX MECS (S-H)/(S+H) (S=1.3-4.5keV, H=4.5-1 keV 
count rates) versus the redshift for
the identified sources. Open circles = broad line, quasars and Sy1; 
filled circles = type 1.8-1.9-2.0
AGN; filled squares = starburst galaxies and LINERS; open triangles =
radio-loud AGN; open squares = optically `normal' galaxies. Dotted lines
show the expected softness ratio for a power law model with
$\alpha_E$=0.4 (lower line) and $\alpha_E=0.8$ (upper line).  Dashed
lines show the expectations of absorbed power law models (with
$\alpha_E=0.8$ and log$N_H$=23.7, 23.0, 22.7, from bottom to top) with
the absorber at the source redshift.
b) Chandra ACIS (S-H)/(S+H) (S=0.5-2keV, H=2-10keV fluxes) versus the 
redshift for the Chandra HELLAS sources (big symbols) and the Barger 
et al. (2000) and
Giacconi et al. (2000) identified sources (small symbols). Dashed
lines show the expectations of absorbed power law models (with
$\alpha_E=0.8$ and log$N_H$=23, 22.7, 22.3, 22, 21.7, 21 and 20.5 
(from bottom to top) with the absorber at the source redshift.
}
\end{figure}

\subsection{Absorption versus extintion in high redshift AGN}

The situation is different at high redshift.  Several broad line AGN
at z$>0.5$ have (S-H)/(S+H) inconsistent with that expected for a
$\alpha_E\sim0.8$ power law. The (S-H)/(S+H) of broad line AGNs in
figure 1a (24) and 2b (18) are marginally anticorrelated with z
(Spearman rank correlation coefficent of -0.364 for 22 dof, and -0.52
for 16 dof, corresponding to probabilities of 92\% and 97.3 \%
respectively).  The number of sources is not large enough to reach a
definite conclusion, but it is interesting to note that this
correlation goes in the opposite direction than expected. In fact, the
ratio of the optical depth in the optical band, due to dust
extinction, to that in the X-ray band, due to photoelectric
absorption, should scale as $(1+z)^4$. Highly X-ray obscured broad
line blue continuum quasar can exist only if their dust to gas ratio
or their dust composition strongly differs from the Galactic one (see
e.g. Maiolino et al, 2000), or if the X-ray absorber is within the
Broad Line Region. Similar results have been recently found in ASCA
samples by Akiyama et al. (2000) and Della Ceca et al. (1999).  A low
dust-to-gas ratio may be obtained if most of the X-ray absorbing gas
is within the sublimation radius of the dust, i.e. close to the
central X-ray source. The sublimation radius depends on the luminosity
of the X-ray source, and so sources with gas at similar distances from
the X-ray source will show low/high extinction depending on the
high/low luminosity of the central source.  Large dust masses
illuminated and heated-up by the strong AGN UV-to-X-ray continuum
would re-emit in the infrared. Based on the average AGN spectral
energy distribution, on the XRB intensity and on the assumption that
most of it is due to obscured active nuclei Fabian \& Iwasawa (1999)
estimated that between 10\% and 50\% of the 100 micron background
could be due to reprocessing of accretion radiation, compared to dust
reprocessed starlight radiation. If however a significant fraction of
the XRB is actually due to broad-line, dust-free, X-ray obscured
quasars, then the contribution to the IR background of the sources
making the hard XRB would be smaller than the above values.

\subsection{X-ray loud optically dull galaxies}

The unprecendented Chandra arcsec position determination allows for
the first time to unambiguously identify X-ray emitting, optically
faint galaxies (Barger et al. 2000, Hornschemeier et al. 2000, Fiore
et al. 2000b, Giacconi this meeting).  In several cases their X-ray to
optical ratio is 10-100 times that of nearby galaxies.  Intriguingly,
the X-ray spectrum of these galaxies is often hard: the softness ratio
in figure 1b implies in several cases substantial column densities,
for a typical AGN X--ray continuum (see figure 1b, open squares).  The
relatively high X-ray luminosity, compared to the optical one,
suggests an X-ray active nucleus in these objects, in which either
continuum beaming dominates (as in BL Lacertae objects), or emission
lines are obscured or not efficiently produced, although more exotic
possibilities (like Advection Dominated Accretion Flow) cannot be
excluded at this point.  AGN optical emission lines could be hidden by
the X-ray absorbing gas.  There are several cases in which X-ray AGN
do not show up in the optical band, see e.g. the highly obscured AGN
NGC4945 (Marconi et al. 2000, Guainazzi et al. 2000) and NGC6240
(Vignati et al. 1999 and references therein). In these cases strong
emission lines are present in the optical spectra, but they are due to
starburst regions, as indicated by their intensity ratios. Conversely,
the X-ray loud optically dull galaxies have very red optical spectra
implying an old stellar population and little star-formation.

\bigskip\noindent
This research has been partly supported by ASI 
ARS/99/75 contract and MURST Cofin-98-032 contract.

%INDEX%%%%%%%%%%%%%%%%%%%%%%%%%%%%%%%%%%%%%%%%%%%%%%%%%%%%%%%%%%%%%%%
\clearpage
\addcontentsline{toc}{section}{Index}
\flushbottom
\printindex
%%%%%%%%%%%%%%%%%%%%%%%%%%%%%%%%%%%%%%%%%%%%%%%%%%%%%%%%%%%%%%%%%%%%%


\begin{thebibliography}{7}
%
\addcontentsline{toc}{section}{References}

\bibitem{}  Akiyama et. al. (2000) Astroph. J, {\bf 532}, 700

\bibitem{} Barger, A., Cowie, L., Mushotzky, R.F., Richards, E.A., (2000),
 Astroph. J. in press, astro-ph/0007175

\bibitem{} Cappi, M. et al. (2000), Astroph. J. in press, astro-ph/0009199


\bibitem{} Comastri, A., Setti, G., Zamorani, G. \& Hasinger, G. (1995), 
Astron. Astroph., {\bf 296}, 1

\bibitem{} Della Ceca, R., Braito, V., Cagnoni, I., Maccacaro, T.
(1999) to appear on Astrophysical Letters and Communications, 
astro-ph/9912016

\bibitem{} Fabian, A., Iwasawa, K. (1999), Montly Notices of the Royal
Astron. Soc., {\bf 303}, L34

\bibitem{} Fabian, A. (1999), Montly Notices of the Royal
Astron. Soc., {\bf 308}, L39


\bibitem{} Fiore, F. et al. (1999), Montly Notices of the Royal
Astron. Soc., {\bf 306}, L55

\bibitem{} Fiore, F. et al. (2000a), Montly Notices of the Royal
Astron. Soc., submitted

\bibitem{} Fiore, F. et al. (2000b), New Astronomy, {\bf 5}, 143, 
astro-ph/0003273


\bibitem{} Giacconi, R. et al. (2000) Astroph. J., submitted,
astro-ph/0007240

\bibitem{} Guainazzi et al. (2000), Astron. Astroph., {\bf 356}, 463 

\bibitem{} Hasinger, G., Lehmann, I., Giacconi, R.,
Schmidt, M. Tr\"umper, J., Zamorani, G., (1999), in 
``Highlights in X-ray Astronomy'' astro-ph/9901103

\bibitem{} Hasinger, G. et al. (2000) Astron. Astroph, in press,
astro-ph/0011271

\bibitem{} Hornschemeier, A.E. et al. (2000), Astroph. J. {\bf 541}, 49

\bibitem{} Marconi, A. et al. (2000) Astron. Astroph., {\bf 357}, 24

\bibitem{} Maiolino, R., Marconi, A., Oliva, E. (2000) Astron. Astroph.,
in press,  astro-ph/0010066

\bibitem{} Vignati et al. (1999), Astron. Astroph, 349, L57


\end{thebibliography}
\end{document}